\documentclass[conference]{IEEEtran}

\usepackage[T1]{fontenc}
\usepackage[bookmarks=false,draft]{hyperref}
\usepackage[cmex10]{amsmath}
\usepackage{amssymb}
\usepackage{amsthm}
\usepackage{amsfonts}
\usepackage{setspace}
\usepackage{fancyhdr}
\usepackage{extramarks}
\usepackage{chngpage}
\usepackage[usenames,dvipsnames]{color}
\usepackage{ifthen}
\usepackage{courier}
\usepackage{graphicx}
\usepackage{textcomp}
\usepackage{amssymb}
\usepackage{cancel}
\usepackage{cases}

\usepackage{cite}

\usepackage{graphicx}
\usepackage{pstricks}

\usepackage{relsize}
\usepackage[font={small}]{caption}

\usepackage{mathtools}
\usepackage{enumitem}

\makeatletter
\def\thm@space@setup{\thm@preskip=2pt
	\thm@postskip=2pt \itshape}
\makeatother

\newtheoremstyle{newstyle}      
{} 
{} 
{\mdseries} 
{} 
{\bfseries} 
{.} 
{ } 
{} 

\theoremstyle{newstyle}

\usepackage[margin=0.75in]{geometry}

\newtheorem{theorem}{Theorem}

\theoremstyle{definition}

\theoremstyle{remark}
\newtheorem{remark}{Remark}

\IEEEoverridecommandlockouts
\begin{document}
	\sloppy
	
	\title{\vspace{0.1in}Coded Fourier Transform}
	\author{Qian~Yu$^{*}$, Mohammad~Ali~Maddah-Ali$^{\dagger}$, and A.~Salman~Avestimehr$^{*}$\\
		$^{*}$ Department of Electrical Engineering, University of Southern California ~~ 
		$^{\dagger}$ Nokia Bell Labs~~~
	}
	\maketitle

	\begin{abstract}
	We consider the problem of computing the Fourier transform of high-dimensional vectors,  distributedly over a cluster of machines consisting of a master node and multiple worker nodes, where the worker nodes can only store and process a fraction of the inputs.  
	We show that by exploiting the algebraic structure of the Fourier transform operation  and leveraging concepts from coding theory, one can efficiently deal with the straggler effects. In particular, we propose a computation strategy, named as \emph{coded FFT}, which achieves the  optimal  recovery threshold, defined as the minimum number of workers that the master node needs to wait for in order to compute the output. 
	This is the first code that achieves the optimum robustness in terms of tolerating stragglers or failures for computing Fourier transforms. Furthermore, the reconstruction process for coded FFT can be mapped to MDS decoding, which can be solved efficiently.  
	Moreover, we extend coded FFT to settings including computing general $n$-dimensional
	Fourier transforms, and provide the optimal computing strategy for those settings. 
	

	\end{abstract}

	\section{Introduction}
 
 Discrete Fourier transform (DFT) is one of the fundamental operations, which has been broadly used in many applications, including signal processing, data analysis, and machine learning algorithms. 
 Due to the increasing size and dimension of data, many modern applications 
 require massive amount of computation and storage, which can not be provided by a single machine. Thus, finding efficient design of algorithms including DFT in
 a distributed computing environment has gained considerable attention. For example, several distributed DFT implementations, such as FFTW \cite{FFTW05} and PFFT \cite{pippig2013pfft}, have been introduced and used widely.   
 
 
 A major performance bottleneck in distributed computing problems is the latency caused by ``stragglers" \cite{dean2013tail}, which are the small fraction of computing nodes at the high latency tail that prolongs the computation.  Mitigating this effect involves creating certain types of ``computation reduncancy'', such that the computation can be completed even without collecting the intermediate results assigned to the stragglers. For example,  one can \emph{replicate} the same computing task onto multiple nodes to provide this redundancy \cite{zaharia2008improving}. 
 
  Recently, it has been shown that \emph{coding theoretic} concepts that were originally developed for communication systems can also be useful in distributed computing systems, playing a transformational role by improving the performance of computation in various aspects. 
  In this context, two ``coded computing'' concepts has been proposed: 
   The first one, introduced in~\cite{LMA_all,li2016fundamental, 7996730}, injects computation redundancy in order to alleviate the communication bottleneck and accelerate distributed computing algorithms (e.g., Coded Terasort~\cite{CTSPaper}). 
The second coded computing concept, introduced in~\cite{lee2015speeding, lee2017high}, 
 utilizes coding to 
 handle the straggler effects and speed up the computations for distributed matrix multiplication. This technique has been further extended 
 to decentralized ``master-less'' architectures~\cite{li2016unified}, distributed convolution \cite{coded_conv17}, short dot linear transform \cite{dutta2016short} and gradient computation \cite{tandon2016gradient}. 

 More recently, \emph{polynomial code}\cite{yu2017polynomial} has been proposed for distributed massive matrix multiplication, for optimal  straggler effect mitigation. It was shown that by designing a pair of codes, whose multiplicative product forms an Maximum Distance Separable (MDS) code, one can orderwise improve upon the prior arts in terms of the \emph{recovery threshold} (i.e., the number of workers that the master needs to wait in order to be able to compute the final output), while optimizing other metrics including computation latency and communication load. This provides the first code that achieves the optimum recovery threshold. 
 Furthermore, it allows mapping the reconstruction problem of the final output to polynomial interpolation, which can be solved efficiently, bridging the rich literature of algebraic coding and distributed matrix multiplication. 
 Moreover, a variation of the polynomial code was applied to coded convolution, and its order-optimality has been proved. 

  
 \begin{figure}[htbp]
			\centering
			\includegraphics[width=0.95\linewidth]{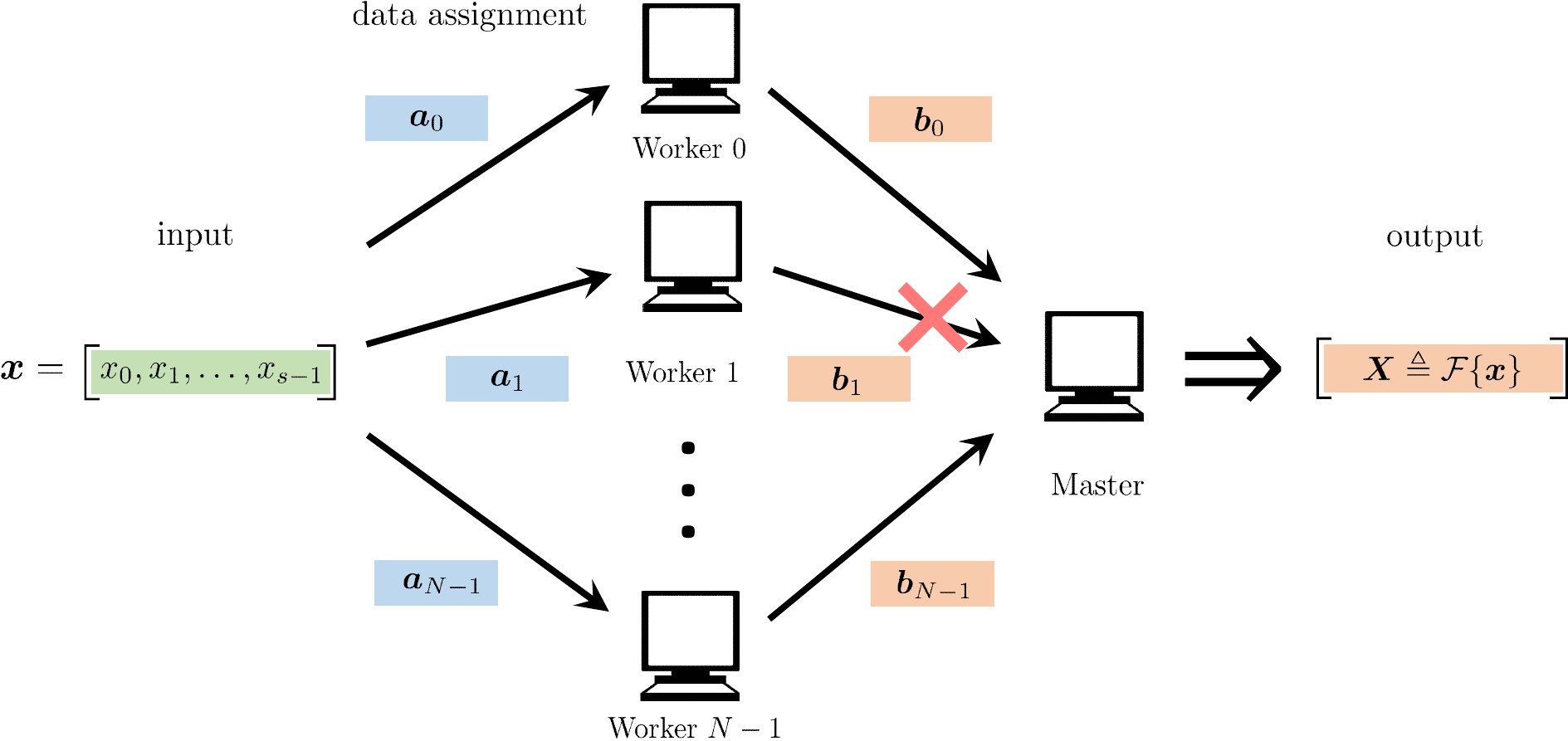}
			\caption{Overview of the distributed Fourier transform framework. Coded data are initially stored  distributedly at $N$ workers according to data assignment. Each worker computes an intermediate result based on its stored vector and returns it to the master. By designing the computation strategy, the master can decode given the computing results from a subset of workers, without having to wait for the stragglers (worker $1$ in this example). }
			\label{fig:sys}
		\end{figure}


In this work, our focus is on mitigating the straggler effects for distributed DFT algorithms.  Specifically, we consider a distributed Fourier transform problem where we aim to compute the discrete Fourier transform
 $\boldsymbol{X}=\mathcal{F}\{\boldsymbol{x}\}$ given an input vector $\boldsymbol{x}$. As shown in Figure \ref{fig:sys}, the computation is carried out using a distributed system with a master node and $N$ worker nodes that can each store and process $\frac{1}{m}$ fraction of the input vector, for some parameter $m\in\mathbb{N}^*$. The vector stored at each worker can be designed as an arbitrary function of the input vector $\boldsymbol{x}$. Each worker can also compute an intermediate result of the same length based on an arbitrary function of its stored vector, and return it to the master. 
By designing the computation strategy at each worker (i.e., designing the functions to store the vector and to compute the intermediate result), the master only need to wait for the fastest subset of workers before recovering the final output $\boldsymbol{X}$, which mitigates the straggler effects. 

Our main result in this paper is the development of an optimal computing strategy, referred to as the \emph{coded FFT}. This computing design achieves the optimum recovery threshold $m$, while allowing the the master to decode the final output with low complexity. Furthermore, we extend this technique to settings including computing multi-dimensional Fourier transform, and propose the corresponding optimal computation strategies. 

To develop coded FFT, we leverage two key algebraic properties of the Fourier transform operations. First due to its recursive structures, we can decompose the DFT into multiple identical and simpler operations (i.e., DFT over shorter vectors), which suits the distributed computing framework and can be potentially assigned to multiple worker nodes. Secondly, due to the linearity of Fourier transform, we can apply linear codes on the input data, which commutes with the DFT operation and translates to the computing results. These two properties allow us to develop a coded computing strategy where the outputs from the worker nodes has certain MDS properties, which can optimally mitigate straggler effects.  



 
 

 	\section{System Model 
 	and Main Results}
 		
 			We consider a problem of computing the Discrete Fouier transform $\boldsymbol{X}=\mathcal{F}\{\boldsymbol{x}\}$ in a distributed computing environment with a master node and $N$ worker nodes. The input $\boldsymbol{x}$ and the output $\boldsymbol{X}$ are vectors of length $s$ over an arbitrary field $\mathbb{F}$ with a primitive $s$th root of unity, denoted by $\omega_s$.\footnote{When the base field  $\mathbb{F}$ is finite, we assume it is sufficiently large.} 
 			We want to compute the elements of the output vector, denoted by $X_0,...,X_{s-1}$, as a function of the elements of the input vector, denoted by $x_0,...,x_{s-1}$, based on the following equations.
 				\begin{align}
 		X_i\triangleq \sum_{j=0}^{s-1}  x_j\omega_s^{ij} &&\textup{for } i\in\{0,\dots,s-1\}.
 		\end{align}
 		
 		Each one of the $N$ workers can store and process $\frac{1}{m}$ fraction of the vector. Specifically,  
 		given a parameter $m\in \mathbb{N}^*$ satisfying $m|s$, 
 			each worker $i$ can store an arbitrary vector $\boldsymbol{a}_i\in \mathbb{F}^{\frac{s}{m}}$ as a function of the input $\boldsymbol{x}$, compute an intermediate result $\boldsymbol{b}_i\in \mathbb{F}^{\frac{s}{m}}$ as a function of $\boldsymbol{a}_i$, and return $\boldsymbol{b}_i$ to the server. The server only waits for the results from a subset of workers, before recovering the final output $\boldsymbol{X}$ using certain \emph{decoding functions}, given these intermediate results returned from the workers.
 			

 		 Given the above system model, we can design the functions to compute $\boldsymbol{a}_i$s' and  $\boldsymbol{b}_i$s' for the workers. We refer to these functions as the \emph{encoding functions} and the \emph{computing functions}.
 		We say that a \emph{computation strategy} consists of $N$ {encoding functions} and $N$ {computing functions}, denoted by  
    	\begin{align}
    	    {\boldsymbol{f}}=(f_0,{f}_1,...,{f}_{N-1}), 
    	\end{align}
    	and 
    		\begin{align}
    	    {\boldsymbol{g}}=(g_0,{g}_1,...,{g}_{N-1}), 
    	\end{align}
    	 that are used to compute the $\boldsymbol{a}_i$s' and  $\boldsymbol{b}_i$s'.
     Specifically, given a computation strategy, each worker $i$ stores $\boldsymbol{a}_i$ and computes  $\boldsymbol{b}_i$ according to the following equations:
    	\begin{align}
    	\boldsymbol{a}_i&={f}_i( \boldsymbol{x} ), \\
    	\boldsymbol{b}_i&={g}_i( \boldsymbol{a}_i ).
    	\end{align}
    	For any integer $k$, we say a computation strategy is \emph{$k$-recoverable} if the master can recover $ X $ given the computing results from \emph{any} $k$ workers using certain decoding functions. We define the \emph{recovery threshold} of a \emph{computation strategy} as the minimum integer $k$ such that the computation strategy 
    	is $k$-recoverable. 
    
	    The goal of this paper is to find the optimal computation strategy that achieves the minimum possible recovery threshold,  while allowing efficient decoding at the master node. 
	    This essentially provides the computation strategy with the maximum robustness against the straggler effect, which only requires a low additional computation overhead.

 We summarize our main results in the following theorems: 

 \begin{theorem}
 \label{thm}
 In a distributed Fourier transform problem of computing $\boldsymbol{X}=\mathcal{F}\{\boldsymbol{x}\}$ using $N$ workers that each can store and process $\frac{1}{m}$ fraction of the input $\boldsymbol{x}$, we can achieve the following recovery threshold
 \begin{align}
 K^*=m.
 \end{align}
 Furthermore, the above recovery threshold can be achieved by a computation strategy, referred to as the \emph{Coded FFT}, which allows efficient decoding at the master node, {i.e.}, with a complexity that scales linearly with respect to the size $s$ of the input data. 
 \end{theorem}

  Moreover, we can prove the optimally of coded FFT, which is formally stated in the following theorem 
  
  \begin{theorem}
 \label{thm:conv}
 In a distributed Fourier transform environment with $N$ workers that each can store and process $\frac{1}{m}$ fraction of the input vector, the following recovery threshold
 \begin{align}
 K^*=m
 \end{align}
 is optimal when the base field $\mathbb{F}$ is finite.\footnote{Similar results can be generalized to the case where the base field is infinite, by taking into account of some practical implementation constrains (see Section \ref{sec:opt}).}
 \end{theorem}
 
  \begin{remark}
  The above 
  converse demonstrates that our proposed coded FFT design is optimal in terms of recovery threshold.
 Moreover, we can prove that coded FFT is also optimal in terms of the communication load (see Section \ref{sec:opt}). 
 \end{remark}

 \begin{remark}
 While in the above results we focused on the developing the optimal coding technique for the one dimensional Fourier transform. The techniques developed in this paper can be easily generalized to the  $n$-dimensional Fourier transform operations. Specifically, we can show that in a general $n$-dimensional Fourier transform setting, the optimum recovery threshold $K^*=m$ can still be achieved, using a generalized version of the coded FFT strategy (see Section \ref{app:a}). Similarly, this also generalized to the scenario where we aim to compute the Fourier transform of multiple input vectors. The optimum recovery threshold $K^*=m$ can also be achieved (see Section \ref{app:b}).

 \end{remark}

 \begin{remark}
 Although the coded FFT strategy was designed focusing on optimally handling the stragglers issues, it can also be applied to the fault tolerance computing setting (e.g., as considered in \cite{4606,293265}, where a module can produce arbitrary error results under failure), to improve robustness to failures in computing.
Specifically, given that the coded FFT produces computing results that are coded by an MDS code, it also enables detecting, or correcting maximum amounts errors even when the erroneous workers can produce arbitrary computing results.
 \end{remark}

 		\section{Coded FFT: the Optimal Computation Strategy}
 		
 		In this section, we prove Theorem \ref{thm} by proposing an optimal computation strategy, referred to as \emph{Coded FFT}. We start by demonstrate this computation strategy and the corresponding decoding procedures
 		through a motivating example.

 		\subsection{Motivating Example}
 		Consider a distributed Fourier transform problem with an input vector $\boldsymbol{x}=[x_0,x_1,x_2,x_3]\in\mathbb{C}^4$, $N=4$ workers, and a design parameter $m=2$. We want to compute the Fourier transform $\boldsymbol{X}=\mathcal{F}\{\boldsymbol{x}\}$, which is specified as follows.
 			\begin{align}\label{eq:fore}
 	\begin{bmatrix}
    X_0   \\   X_1   \\
    X_2  \\   X_3 
\end{bmatrix}
&=
	    \begin{bmatrix}
    1 &1&1&1 \\
   1&-\sqrt{-1}&-1&\sqrt{-1} \\
   1&-1&1&-1 \\
   1&\sqrt{-1}&-1&-\sqrt{-1} 
\end{bmatrix}
 	\begin{bmatrix}
     x _0   \\    x_1\\
     x _2    \\    x_3
\end{bmatrix}.
	\end{align}	
 		We aim to design a computation strategy to achieve a recovery threshold of $2$. 
 		

 			\begin{figure*}[htbp]
			\centering
			\includegraphics[width=0.9\linewidth]{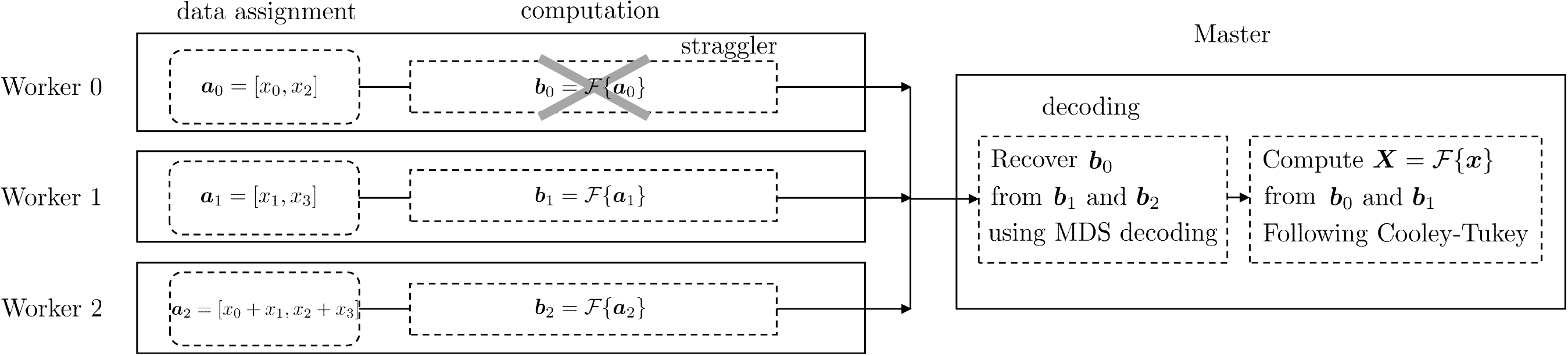}
			\caption{Example using coded FFT, with $3$ workers that can each store and process half of the input. (a) Computation strategy: each worker $i$ stores a linear combination of the interleaved version of the input according to an MDS code, and computes its DFT. (b) Decoding: master waits for results from \emph{any} $2$ workers, and recover the final output by first decoding the MDS code, then compute the transformed vector following the similar steps in the Cooley-Tukey algorithm. 
			}
			\label{fig:exp}
		\end{figure*}
 		
 	
 	In order to design the optimal strategy, we exploit two key properties of the DFT operation.
 			Firstly, DFT has the following recursive structure: 
 			\begin{align}
 		    X_{i}&=\sum_{j=0}^{3}  x_j(-\sqrt{-1})^{ij}\\
 		    &=\sum_{k=0}^{1} c_{0,k} {(-1)}^{ik}+(-\sqrt{-1})^{i}\sum_{k=0}^{1} c_{1,k} {(-1)}^{ik},
 		\end{align}
 		where vectors $\boldsymbol{c}_0$ and $\boldsymbol{c}_1$ are the interleaved version of the input vector:
 			\begin{align}
 		\boldsymbol{c}_0&=[x_0,x_2],\\
 		\boldsymbol{c}_1&=[x_1,x_3].
 		\end{align}
        This structure decomposes the Fourier transform into two identical and simpler operations: the Fourier transform of $\boldsymbol{c}_0$ and $\boldsymbol{c}_1$, defined as follows.
        \begin{align}
        {C}_{i,j}\triangleq \sum_{k=0}^{1} c_{i,k} {(-1)}^{jk}.
        \end{align}
        Hence, computing the Fourier transform of a vector is essentially computing the Fourier transforms of its sub-components. This property has been exploited in the context of single machine algorithms and led to the famous Cooley-Tukey algorithm \cite{cooley1965algorithm}.
        
        On the other hand, we exploit the linearity of the DFT operation to inject linear codes in the computation to provide robustness against stragglers. Specifically, given that the Fourier transform of any linearly coded vector equals the linear combination of the Fourier transforms of the individual vectors, by injecting MDS code on the interleaved vectors $\boldsymbol{c}_0$ and $\boldsymbol{c}_1$ and computing their Fourier transforms, we obtain a coded version of the vectors $\boldsymbol{C}_0$ and $\boldsymbol{C}_1$. This provides the redundancy to mitigate the straggler effects.
        
        Specifically, we encode $\boldsymbol{c}_0$ and $\boldsymbol{c}_1$ using a $(3,2)$-MDS code, and let each worker store one of the coded vectors. I.e., 
 		\begin{align}
 		\boldsymbol{a}_0&=\boldsymbol{c}_0, \\
 		\boldsymbol{a}_1&=\boldsymbol{c}_1,\\
 		\boldsymbol{a}_2&=\boldsymbol{c}_0+\boldsymbol{c}_1.
 		\end{align}
 		Each worker computes the Fourier transform $\boldsymbol{b}_i=\mathcal{F}{\{\boldsymbol{a}_i\}}$ of its assigned vector. Specifically, each worker $i$ computes
 		\begin{align}
 	\begin{bmatrix}
    b_{i,0}   \\   b_{i,1}  
\end{bmatrix}
&=
	    \begin{bmatrix}
    1 &1 \\
   1&-1
\end{bmatrix}
 	\begin{bmatrix}
     a _{i,0}   \\    a_{i,1}
\end{bmatrix}.
	\end{align}	
	To prove that this computation strategy gives a recovery threshold of $2$, we need to design a valid decoding function for any subset of $2$
workers. We demonstrate this decodability through a representative scenario, where the master receives the computation results
from worker $1$ and worker $2$ as shown in Figure \ref{fig:exp}. The decodability for the other $2$ possible scenarios can be proved similarly.

According to the designed computation strategy, the server can first recover the computing result of worker $0$ given the results from the other workers as follows:
\begin{align}
    \boldsymbol{b_0}=\boldsymbol{b}_2-\boldsymbol{b}_1.
\end{align}
After recovering $\boldsymbol{b}_0$, we can verify that the server can then recover the final output $\boldsymbol{X}$ using $\boldsymbol{b}_0$ and $\boldsymbol{b}_1$ as follows:
			\begin{align}
 	\begin{bmatrix}
    X_0   \\   X_1   \\
    X_2  \\   X_3 
\end{bmatrix}
=
 	\begin{bmatrix}
     b_{0,0}+ b_{1,0}   \\    b_{0,1} -\sqrt{-1} \cdot b_{1,1}\\
    b_{0,0} - b_{1,0}   \\    b_{0,1} + \sqrt{-1}\cdot b_{1,1}
\end{bmatrix}.
	\end{align}	
 		
 		\subsection{General Description of Coded FFT}
 		
 		Now we present an optimal computing strategy that achieves the optimum recovery threshold stated in Theorem \ref{thm}, for any parameter values of $N$ and $m$. First of all we interleave the input vector $\boldsymbol{x}$ into $m$ vectors of length $\frac{s}{m}$, denoted by $\boldsymbol{c}_0,...,\boldsymbol{c}_{m-1}$. Specifically, we let the $j$th element of each $\boldsymbol{c}_i$ equal
 		\begin{align}
 		    c_{i,j}=x_{i+jm}.
 		\end{align}
 		We denote the discrete Fourier transform of each interleaved vector $\boldsymbol{c}_i$, in the domain of $\mathbb{Z}_{\frac{s}{m}}$, as $\boldsymbol{C}_i$. Specifically,
 			\begin{align}
 		C_{i,j}\triangleq \sum_{k=0}^{\frac{s}{m}-1}  c_{i,k}\omega_s^{jkm} &&\textup{for } j\in\{0,\dots,\frac{s}{m}-1\}.
 		\end{align}
 		Note that if the master node can recover all the above Fourier transform $\boldsymbol{C}_i$ of the interleaved vectors, 
 		the final output can be computed based on the following identities:
 			\begin{align}
 		    X_{i}&=\sum_{j=0}^{m-1}\sum_{k=0}^{\frac{s}{m}-1} c_{j,k} \omega_s^{i(j+km)}\\
 		    &=\sum_{j=0}^{m-1} C_{j,\bmod (i,\frac{s}{m})} \omega_s^{ij},\label{eq:sf}
 		\end{align}
 		where $\bmod (i,\frac{s}{m})$ denotes the remainder of $i$ divided by $\frac{s}{m}$.

 	    Based on this observation, we can naturally view the distributed Fourier transform problem as a problem of distributedly computing a list of linear transformations, i.e., computing the Fourier transform of $\boldsymbol{c}_i$'s. We inject the redundancy as follows to provide robustness to the computation:
 	
 		 We first encode the $\boldsymbol{c}_0, \boldsymbol{c}_1, ..., \boldsymbol{c}_{m-1}$ using an arbitrary $(N,m)$-MDS code, where the coded vectors are denoted $\boldsymbol{a}_0,...,\boldsymbol{a}_{N-1}$ and are assigned to the workers correspondingly. 
 		Then each worker $i$ computes  the Fourier of $\boldsymbol{a}_i$, and return it to the master. Given the linearity of Fourier transform, the computing results $\boldsymbol{b}_0,...,\boldsymbol{b}_{N-1}$ are essentially linear combinations of the Fourier transform $\boldsymbol{C}_i$'s, which are coded by the same MDS code. Hence, after the master receives any $m$ computing results, it can decode the message $\boldsymbol{C}_i$'s, and proceed to recover the final result. This allows achieving the recovery threshold of $m$.

 		 \begin{remark}
 The recovery threshold $K^*=m$ achieved by coded FFT can not be achieved using computation strategies that were developed for generic matrix-by-vector multiplication in the literature \cite{lee2015speeding,dutta2016short}. Specifically, the conventional uncoded repetition strategy requires a recovery threshold of $N-\frac{N}{m^2}+1$, and the short-dot (or short-MDS) strategy provided in \cite{lee2015speeding,dutta2016short} requires $N-\frac{N}{m}+m$. 
 Hence, by developing a coding strategy for the specific purpose of computing Fourier transform, we can achieve order-wise  improvement in the recovery threshold.
 \end{remark}
 	
 		\subsection{Decoding Complexity of Coded FFT}
 	
 	Now we show that coded FFT allows an efficient decoding algorithm at the master for recovering the output. 
 	After receiving the computing results, the master needs to recover the output in two steps: decoding the MDS code and then computing $\boldsymbol{X}$ from the intermediate value $\boldsymbol{C}_i$'s.
 	
 	For the first step, the master needs of decode an $(N,m)$-MDS code by $\frac{s}{m}$ times. This can be computed efficiently, by selecting an MDS code with low decoding complexity for the coded FFT design. There has been various works on finding efficiently decodable MDS codes (e.g.,\cite{didier2009efficient,5421749}). In general, an upper bound on the decoding complexity of $(N,m)$-MDS code is given by $O(m\log^2 m \log\log m)$, which can be attained by the \emph{Reed-Solomon} codes \cite{roth2006introduction} and using fast polynomial interpolation \cite{kedlaya2011fast} as the decoding algorithm. Consequently, the first step of the decoding algorithm has a complexity of at most $O(s\log^2 m \log\log m)$, which scales linearly with respect to $s$.
 	
 	For the second step, the master node needs to evaluate equation (\ref{eq:sf}) to recover the final result. Equivalently, the master needs to compute
 		\begin{align}
 		    X_{i+j\frac{s}{m}}&=\sum_{k=0}^{m-1} C_{k,i} \omega_s^{ik+j  k \frac{s}{m}}
 		\end{align}
 	for any $i\in\{0,1,...,\frac{s}{m}-1\}$ and $j\in\{0,...,m-1\}$. This is essentially the Fourier transform of $\frac{s}{m}$ vectors of length $m$, where the $k$th element of the $i$th vector equals  $C_{k,i} \omega_s^{ik}$. In most cases (e.g., $\mathbb{F}=\mathbb{C}$), the Fourier transform of a length $m$ vector can be efficiently computed with a complexity of $O(m\log m)$, which is faster than the corresponding MDS decoding procedure used in the first step. In general, the computational complexity of Fourier transform is upper bounded by  $O(m\log m \log\log m)$, which can be achieved by a combination of Bluestein's algorithm and fast polynomial multiplication \cite{cantor1991fast}. Hence, the complexity of the second step is at most $O(s\log m \log\log m)$.
 	
 	To conclude, 
 	 	our proposed coded FFT strategy allows efficient decoding with a complexity of at most $O(s\log^2 m \log\log m)$, which is linear to the input size $s$. The decoding computation is bottlenecked by the first step of the algorithm, which is essentially decoding an $(N,m)$-MDS code by $\frac{s}{m}$ times. To achieve the best performance, one can pick any MDS code with a decoding algorithm that requires the minimum amount of computation based on the problem scenatio \cite{Baktir:2006:AEP:1185448.1185568}.

  	\section{Optimality of coded FFT}
 			\label{sec:opt}
 			In this section, we prove Theorem \ref{thm:conv} through a matching information theoretic converse. 
Specifically, we need to prove that for any computation strategy, the master needs to wait for at least $m$ workers in order to recover the final output. 

Recall that Theorem \ref{thm:conv} is stated for finite fields, we can let the input $\boldsymbol{x}$ be 
be uniformly randomly sampled from $\mathbb{F}^{s}$. Given the invertibility of the Discrete Fourier transform, the output vector $\boldsymbol{X}$ given this input distribution must also be uniformly random on $\mathbb{F}^{s}$. This means that the master node essentially needs to recover a random variable with entropy of $H(\boldsymbol{X})=s\log_2 |\mathbb{F}|$ bits. Note that each worker returns $\frac{s}{m}$
elements of $\mathbb{F}$, providing at most $\frac{s}{m}\log_2 |\mathbb{F}|$ bits of information.
By applying a cut-set bound around the master, we can show that at least results from $m$ workers need to be
collected. Thus we have that the recovery threshold $K^*=m$ is optimal.

 			\begin{remark}
 			Besides the recovery threshold, communication load is also an important metric in distributed computing. The above cut-set converse in fact directly bounds the needed communication load for computing Fourier transform directly, proving that at least $s\log_2 |\mathbb{F}|$ bits of communication is needed. Note that our proposed coded FFT uses exactly this amount of communication to deliver the intermediate results to the server. Hence, it is also optimal in terms of communication.   
 			\end{remark}
 			
 			\begin{remark}
 			Although Theorem \ref{thm:conv} focuses on the scenario where the base field $\mathbb{F}$ is finite, 
  similar results can be obtained when  the base field is infinite (e.g., $\mathbb{F}=\mathbb{C}$), by taking into account of the practical implementation constrains. For example, any computing device can only keep variables reliably with finite precision. This quantization requirement in fact allows applying the cut-set bound for the distributed Fourier transform problem, even when $\mathbb{F}$ is infinite, and enables proving the optimally of coded FFT in those scenarios.
 		
 			\end{remark}


\section{$n$-dimensional Coded FFT }\label{app:a}
Fourier transform in higher dimensional spaces is a frequently used operation in image processing and machine learning applications. In this section, we consider the problem of designing optimal codes for this operation. We show that the coded FFT strategy can be  naturally extended to this scenario, and achieves the optimum performances. We start by formulating the system model and state the main results.

\subsection{System Model and Main results}

			We consider a problem of computing an $n$-dimensional Discrete Fourier transform $T=\mathcal{F}\{t\}$ in a distributed computing environment with a master node and $N$ worker nodes. The input $t$ and the output $T$ are tensors of order $n$, with dimension $s_0\times s_1\times...\times s_{n-1}$.  For brevity, we denote the total number of elements in each tensor by $s$, i.e., $s\triangleq s_0s_1...s_{n-1}$.

			The elements of the tensors belong to a field $\mathbb{F}$ with a primitive $s_k$th root of unity for each $k\in\{0,...,n-1\}$, denoted by $\omega_{s_k}$. 
 			We want to compute the elements of the output tensor $T$, denoted by $\{T_{i_0i_1...i_{n-1}}\}_{i_\ell\in\{0,...,s_i-1\}, \forall \ell\in\{0,...,n-1\}}$, as a function of the elements of the input tensor, denoted by $\{t_{i_0i_1...i_{n-1}}\}_{i_\ell\in\{0,...,s_i-1\}, \forall \ell\in\{0,...,n-1\}}$, based on the following equations.
 				\begin{align}
 		T_{i_0i_1...i_{n-1}}\triangleq \sum_{\substack{j_\ell\in\{0,...,s_i-1\}, \\ \forall \ell\in\{0,...,n-1\}}}  t_{j_0j_1...j_{n-1}}\prod_{k=0}^{n-1}\omega_{s_k}^{i_kj_k} .
 		\end{align}

 		Each one of the $N$ workers can store and process $\frac{1}{m}$ fraction of the tensor. Specifically,  
 		given a parameter $m\in \mathbb{N}^*$ satisfying $m|s$, 
 			each worker $i$ can store an arbitrary vector $\boldsymbol{a}_i\in \mathbb{F}^{\frac{s}{m}}$ as a function of the input $\boldsymbol{t}$, compute an intermediate result $\boldsymbol{b}_i\in \mathbb{F}^{\frac{s}{m}}$ as a function of $\boldsymbol{a}_i$, and return $\boldsymbol{b}_i$ to the server. The server only waits for the results from a subset of workers, before recovering the final output $T$ using certain \emph{decoding functions}, given these intermediate results returned from the workers.

 		 Similar to the one dimensional Fourier transform problem, we design the functions to compute $\boldsymbol{a}_i$s' and  $\boldsymbol{b}_i$s' for the workers, and refer to them as the 
 	 \emph{computation strategy}. We aim to find an optimal computation strategy that achieves the minimum possible recovery threshold,  while allowing efficient decoding at the master node.

	    Our main results are summarized in the following theorems: 

 \begin{theorem}
 \label{thm_apa}
 In an $n$-dimensional distributed Fourier transform problem of computing $T=\mathcal{F}\{\boldsymbol{t}\}$ using $N$ workers that each can store and process $\frac{1}{m}$ fraction of the input $t$, we can achieve the following recovery threshold
 \begin{align}
 K^*=m.
 \end{align}
 Furthermore, the above recovery threshold can be achieved by a computation strategy, referred to as the \emph{$n$-dimentional Coded FFT}, which allows efficient decoding at the master node, {i.e.}, with a complexity that scales linearly with respect to the size $s$ of the input data. 
 \end{theorem}
 
  Moreover, we can prove the optimally of $n$-dimensional coded FFT, which is formally stated in the following theorem. 
  
  \begin{theorem}
 \label{thm:conv_apa}
 In an $n$-dimensional distributed Fourier transform environment with $N$ workers that each can store and process $\frac{1}{m}$ fraction of the input vector from a finite field $\mathbb{F}$, the following recovery threshold
 \begin{align}
 K^*=m
 \end{align}
 is optimal.\footnote{Similar to the $1$-dimensional case, this optimally can be generalized to base fields with infinite cardinally, by taking into account of some practical implementation constrains.}
 \end{theorem}
 
 	\subsection{General Description of $n$-dimensional Coded FFT}\label{sec:ndg}
 		
 	We first prove Theorem \ref{thm_apa} by proposing an optimal computation strategy, referred to as \emph{$n$-dimensional Coded FFT}, that achieves the recovery threshold $K^*=m$ for any parameter values of $N$ and $m$. 
 	
	    
	    	First of all we interleave the input tensor $\boldsymbol{t}$ into $m$ smaller tensors, each with a total size of $\frac{s}{m}$. Specifically, given that $m|s$, we can find integers $m_0,m_1,...,m_{n-1}\in \mathbb{N}$, such that $m_k|s_k$ for each $k\in\{0,1,...,n\}$, and for each tuple $(i_0,i_1,...,i_{n-1})$ satisfying $i_k\in\{0,1,...,m_k-1\}$, we define a tensor $c_{i_0i_1...i_{n-1}}$ with dimension $\frac{s_0}{m_0}\times \frac{s_1}{m_1}\times...\times \frac{s_{n-1}}{m_n-1}$, with the following elements: 
 		\begin{align}
 		    c_{i_0i_1,...i_{n-1},j_0j_1,...j_{n-1}}=t_{(i_0+j_0m) (i_1+j_1m) ... (i_{n-1}+j_{n-1}m)}.
 		\end{align}
 		We denote the discrete Fourier transform of each interleaved tensor  $c_{i_0i_1...i_{n-1}}$ by  $C_{i_0i_1...i_{n-1}}$. Specifically,
 			\begin{align}
 		C_{i_0i_1,...i_{n-1},j_0j_1,...j_{n-1}}&\triangleq \\ \sum_{\substack{j'_\ell\in\{0,...,\frac{s_i}{m_i}-1\}, \\ \forall \ell\in\{0,...,n-1\}}}   &c_{i_0i_1,...i_{n-1},j'_0j'_1...j'_{n-1}} \prod_{k=0}^{n-1} \omega_{s_k}^{j_k j'_k m_k}
 		\end{align}
 		for any $j_{\ell} \in\{0,...,\frac{s_i}{m_i}-1\}$. 
 		
 		Note that if the master node can recover all the above Fourier transform $C_{i_0i_1...i_{n-1}}$ of the interleaved tensors, 
 		the final output can be computed based on the following identity:
 			\begin{align}
 		    	T_{i_0i_1...i_{n-1}}&=\sum_{\substack{j_\ell\in\{0,...,m_i-1\}, \\ \forall \ell\in\{0,...,n-1\}}} C_{j_0j_1...j_{n-1}, i'_0i'_1...i'_{n-1}} \prod_{k=0}^{n-1} \omega_{s_k}^{i_k j_k},
 		\end{align}
 		where $i'_\ell={ \bmod (i_\ell,\frac{s_\ell}{m_\ell})}$. 
 	    Hence, we can view this distributed Fourier transform problem as a problem of computing a list of linear transformations, and we inject the redundancy using MDS code similar to the one dimensional coded FFT strategy. 
 	
 		 Specifically, we encode the $c_{i_0i_1...i_{n-1}}$'s using an arbitrary $(N,m)$-MDS code, where the coded tensors are denoted $\boldsymbol{a}_0,...,\boldsymbol{a}_{N-1}$ and are assigned to the workers correspondingly. 
 		Then each worker $i$ computes the Fourier of tensor $\boldsymbol{a}_i$, and return it to the master. Given the linearity of Fourier transform, the computing results $\boldsymbol{b}_0,...,\boldsymbol{b}_{N-1}$ are essentially linear combinations of the Fourier transform $C_{i_0i_1...i_{n-1}}$'s, which are coded by the same MDS code. Hence, after the master receives any $m$ computing results, it can decode the message $C_{i_0i_1...i_{n-1}}$'s, and proceed to recover the final result. This allows achieving the recovery threshold of $m$.
 		
 	In terms of the decoding complexity, $n$-dimensional coded FFT also requires first decoding an MDS code, and then recovering the final result by computing Fourier transforms of tensors with lower dimension. Similar to the one dimensional FFT, the bottleneck of the decoding algorithm is also the first step, which requires decoding an $(N,m)$-MDS code by $\frac{s}{m}$ times. This decoding complexity is upper bounded by $O(s\log^2 m \log\log m)$, which is linear with respect to the input size $s$. It can be further improved in practice by using any MDS code or MDS decoding algorithms with better computational performances.

	\subsection{Optimally of $n$-dimensional Coded FFT}

    The optimally of $n$-dimensional Coded FFT (i.e., Theorem \ref{thm:conv_apa}) can be proved as follows. When the base field $\mathbb{F}$ is finite, let the input $t$ be 
be uniformly randomly sampled from $\mathbb{F}^{s}$. Given the invertibility of the $n$-dimensional Discrete Fourier transform, the output tensor $\boldsymbol{T}$ given this input distribution must also be uniformly random on $\mathbb{F}^{s}$. 
Hence,  the master node needs to collect at least $H(\boldsymbol{T})=s\log_2 |\mathbb{F}|$ bits of information, where each worker can provide at most $\frac{s}{m}\log_2 |\mathbb{F}|$ bits.
By applying the cut-set bound around the master, we can prove that at least $m$ worker needs to return their results to finish the computation.
 			
 Moreover, the above converse can also be extended to prove that the  $n$-dimensional Coded FFT is optimal in terms of communication.

 \section{Coded FFT with multiple inputs}\label{app:b}
 
 Coded FFT can also be extended to optimally handle computation tasks with multiple inputs entries. In this section, we consider the problem of designing optimal codes for such scenario. 

\subsection{System Model and Main results}

			We consider a problem of computing the $n$-dimensional Discrete Fourier transform of $q$ input tensors, in a distributed computing environment with a master node and $N$ worker nodes. The inputs, denoted by $t_0,t_1,...,t_{q-1}$, are $q$ tensors of order $n$ and dimension $s_0\times s_1\times...\times s_{n-1}$.  For brevity, we denote the total number of elements in each tensor by $s$, i.e., $s\triangleq s_0s_1...s_{n-1}$. 
			The elements of the tensors belong to a field $\mathbb{F}$ with a primitive $s_k$th root of unity for each $k\in\{0,...,n-1\}$, denoted by $\omega_{s_k}$.   
 			We aim to compute the Fourier transforms of the input tensors, which are denoted by $T_0,T_1,...,T_{q-1}$. Specifically, we want to compute the
 			elements of the output tensors according to the following equations.
 				\begin{align}
 		T_{h,i_0i_1...i_{n-1}}\triangleq \sum_{\substack{j_\ell\in\{0,...,s_i-1\}, \\ \forall \ell\in\{0,...,n-1\}}}  t_{h,j_0j_1...j_{n-1}}\prod_{k=0}^{n-1}\omega_{s_k}^{i_kj_k} .
 		\end{align}

 		Each one of the $N$ workers can store and process $\frac{1}{m}$ fraction of the entire input. Specifically,  
 		given a parameter $m\in \mathbb{N}^*$ satisfying $m|qs$, 
 			each worker $i$ can store an arbitrary vector $\boldsymbol{a}_i\in \mathbb{F}^{\frac{qs}{m}}$ as a function of the input tensors, compute an intermediate result $\boldsymbol{b}_i\in \mathbb{F}^{\frac{qs}{m}}$ as a function of $\boldsymbol{a}_i$, and return $\boldsymbol{b}_i$ to the server. The server only waits for the results from a subset of workers, before recovering the final output using certain \emph{decoding functions}.

 		For this problem, we can find an optimal computation strategy that achieves the minimum possible recovery threshold,  while allowing efficient decoding at the master node.
	     We summarize this result in the following theorems: 

 \begin{theorem}
 \label{thm_apb}
 For an $n$-dimensional distributed Fourier transform problem using $N$ workers, if each worker can store and process $\frac{1}{m}$ fraction of the $q$ inputs, we can achieve the following recovery threshold
 \begin{align}
 K^*=m.
 \end{align}
 Furthermore, the above recovery threshold can be achieved by a computation strategy, which allows efficient decoding at the master node, {i.e.}, with a complexity that scales linearly with respect to the size $s$ of the input data. 
 \end{theorem}
 
  Moreover, we  prove the optimally of our proposed computation strategy, which is formally stated in the following theorem. 
  
  \begin{theorem}
 \label{thm:conv_apb}
 In an $n$-dimensional distributed Fourier transform environment with $N$ workers that each can store and process $\frac{1}{m}$ fraction of the input vector, the following recovery threshold
 \begin{align}
 K^*=m
 \end{align}
 is optimal when the base field $\mathbb{F}$ is finite.\footnote{Similar to the single input case, this optimally can be generalized to base fields with infinite cardinally, by taking into account of some practical implementation constrains.}
 \end{theorem}
 
 	\subsection{General Description of Coded FFT with Multiple Inputs}
 		
 	We prove Theorem \ref{thm_apb} by proposing an optimal computation strategy that achieves the recovery threshold $K^*=m$. 
 	First of all we interleave the $q$ inputs into smaller tensors. Specifically, given that $m|qs$, we can find integers $\tilde{m},m_0,m_1,...,m_{n-1}\in \mathbb{N}$, such that $\tilde{m}|q$ and $m_k|s_k$ for each $k\in\{0,1,...,n\}$. For each input tensor $t_h$ and each tuple $(i_0,i_1,...,i_{n-1})$ satisfying $i_k\in\{0,1,...,m_k-1\}$, we define a tensor $c_{h, i_0i_1...i_{n-1}}$ with dimension $\frac{s_0}{m_0}\times \frac{s_1}{m_1}\times...\times \frac{s_{n-1}}{m_n-1}$, with the following elements: 
 		\begin{align}
 		    c_{h,i_0i_1,...i_{n-1},j_0j_1,...j_{n-1}}=t_{h,(i_0+j_0m) (i_1+j_1m) ... (i_{n-1}+j_{n-1}m)}.
 		\end{align}
 		
 	As explained in Section \ref{sec:ndg}, if the master node can obtain the Fourier transforms of all the   interleaved tensors, then
 		the final outputs can be computed efficiently. 
 	    Hence, we can view this distributed Fourier transform problem as a problem of computing a list of linear transformations, and we inject the redundancy using MDS code similar to the single input coded FFT strategy. 
 	
 		 Specifically, we first bundle the $q$ input tensors into $\tilde{m}$ disjoint subsets of same size. For convenience, we denote the set of indices for the $i$th subset by $\mathcal{S}_i$. Within each subset, we view all interleaved tensors with the same index parameter $(i_0,i_1,...,i_{n-1})$ as one message symbol and we encode all the symbols using an arbitrary $(N,m)$-MDS code.
 		 More precisely, for each $g\in\{0,1,...,\tilde{m}-1\}$ and each index parameter $(i_0,i_1,...,i_{n-1})$,  we create the following symbol $\{c_{h, i_0i_1...i_{n-1}}\}_{h\in\mathcal{S}_g}$. There are $m$ symbols in total and we encode them using an $(N,m)$-MDS code. We assign the $N$ coded symbols to $N$ workers, and each of them computes the Fourier transform of all coded tensors contained in the symbol. 
 		 
 		  Given the linearity of Fourier transform, the computing results $\boldsymbol{b}_0,...,\boldsymbol{b}_{N-1}$ are essentially linear combinations of the Fourier transforms of the interleaved tensors, which are coded by the same MDS code. Hence, after the master receives any $m$ computing results, it can decode all the needed intermediate values, and proceed to recover the final result. This allows achieving the recovery threshold of $m$.
 		
 	In terms of the decoding complexity, one can show that the bottleneck of the decoding algorithm is the decoding of the $(N,m)$-MDS code by $\frac{s}{m}$ times, using similar arguments mentioned in Section \ref{app:a}. This decoding complexity is upper bounded by $O(s\log^2 m \log\log m)$, which is linear with respect to the input size $s$. It can be further improved in practice by using any MDS code or MDS decoding algorithms with better computational performances.

	\subsection{Optimally of Coded FFT with multiple inputs}

    The optimally of our proposed Coded FFT strategy for multiple users (i.e., Theorem \ref{thm:conv_apb}) can be proved as follows. When the base field $\mathbb{F}$ is finite, let the input tensors be 
be uniformly randomly sampled from $\mathbb{F}^{q\times s}$. Given the invertibility of the Discrete Fourier transform, the output tensors must also be uniformly random on $\mathbb{F}^{q\times s}$. 
Hence,  the master node needs to collect at least $qs\log_2 |\mathbb{F}|$ bits of information, where each worker can provide at most $\frac{qs}{m}\log_2 |\mathbb{F}|$ bits.
By applying the cut-set bound around the master, we can prove that at least $m$ worker needs to return their results to finish the computation.
 			
 Moreover, the above converse also applies for proving the optimally of Coded FFT in terms of communication. 		
	
	\section{Conclusions}
	We considered the problem of computing the Fourier transform of high-dimensional vectors,  distributedly over a cluster of machines.  We propose a computation strategy, named as \emph{coded FFT}, which achieves the  optimal  recovery threshold, defined as the minimum number of workers that the master node needs to wait for in order to compute the output. We also extended coded FFT to settings including computing general $n$-dimensional
	Fourier transforms, and provided the optimal computing strategy for those settings. There are several interesting future directions, including the practical demonstration of coded FFT over distributed clusters, generalization of coded FFT to more general master-less architectures, and extension of coded FFT to other computing architectures (e.g., edge and fog computing architectures~\cite{li2016edge,li2017scalable,LiCommMagazine}).

	\section{Acknowledgement}
This work is in part supported by NSF grant CIF 1703575, ONR award N000141612189, and a research gift from Intel. 
This material is based upon work supported by Defense Advanced Research Projects Agency (DARPA) under Contract No. HR001117C0053. The views, opinions, and/or findings expressed are those of the author(s) and should not be interpreted as representing the official views or policies of the Department of Defense or the U.S. Government.

\bibliographystyle{ieeetr}
\bibliography{fft}
\end{document}